\documentclass[useAMS,usenatbib,onecolumn]{mn2ex2}
\usepackage{graphicx}
\usepackage{graphics}
\usepackage{amssymb}

\title[A Secular Relativistic Model]{A Secular Relativistic Model For Solar System's Numerical Simulations}
\author[T. Gallardo and J. Venturini]{Tabar\'e Gallardo$^{1}$\thanks{E-mail:
gallardo@fisica.edu.uy} and Julia Venturini$^{1}$\thanks{E-mail:
jventurini@fisica.edu.uy} \\
$^{1}$Facultad de Ciencias, Instituto de Fisica, Igua 4225, Montevideo, 11400, Uruguay}

\begin{document}

\pagerange{\pageref{firstpage}--\pageref{lastpage}} \pubyear{2009}

\maketitle

\label{firstpage}

\maketitle

\begin{abstract}

Using Gauss' averaged equations, we compute the secular relativistic effects generated by the Sun on the argument of the perihelion and the
mean anomaly of an orbit. Then we test
different alternative simpler models that have been proposed to reproduce the secular relativistic effects in the orbital elements.
Generally, models introduce artificial perturbations
that are velocity-independent but that depend on the heliocentric distance. If these perturbations are set as an impulse in a constant
timestep integrator, when the particle approaches perihelion the generated
impulse could be very strong and badly sampled, originating a spurious
orbital evolution. In order to overcome this setback,
we propose
two new models based on a constant, distance-independent, perturbation. With these models we
obtain the correct secular drift in the argument of perihelion and the expected secular orbital evolution is reproduced.
We also discuss with some detail the secular effect generated on the mean anomaly by different models.
This work is an expanded version of \citet{vega}.

\end{abstract}

\section{Introduction}

 The problem of computing relativistic effects in planetary systems and in particular in the
Solar System can be modeled by a perturbation to the Newtonian acceleration and is of increasing importance for future study of low
perihelia and low semimajor axis populations. For our planetary system,
the relevant relativistic effects are those generated by the Sun (see for example \citealt{Benitez}).
It is a usual practice in textbooks to show the relativistic effect
in a planet's perihelion starting from a simplified problem that
gives rise to a unique radial perturbation to the Newtonian
potential conserving the angular momentum.  But, a more rigorous
analysis \citep{Brumbergbook,saye94,bertotti} intended to agree with the IAU standards \citep{IAU}
gives rise also to a
transverse component
and
the expression for the relativistic perturbation becomes:

\begin{equation}\label{quinn} \Delta\ddot\textbf{r}=
\frac{\mu}{r^{3}c^{2}}\left[\left(\frac{4\mu}{r}-\textbf{v}^{2}\right)\textbf{r}+4(\textbf{v.r})\textbf{v}\right]
\end{equation}
where $\mu=k^2 m_{\odot}$  and $\textbf{r}$, $\textbf{v}$ are heliocentric \citep{ander75,Quinn}. This expression, that has no normal component, is
valid when considering the relativistic effects generated only by
a spherically symmetric and non rotating
 Sun.
 When the rotation of the Sun is considered a new effect called gravitomagnetic Lense-Thirring effect appears
 \citep{soffel89,iorio05a}.
 Furthermore, if all bodies have relativistic contributions a more
complex expression should be used as explained in, for example, \cite{Benitez}. The time in Eq. (\ref{quinn}) should be considered as the
\textit{coordinate time}; in order to compare with observables it is necessary to transform, for example, to
Barycentric Coordinate Time (TCB) or Terrestrial Time (TT) \citep{saye94,IAU}.

In order to avoid either the computational cost or the non sympletic form of Eq. (\ref{quinn}), some simpler models that only depend on $r$ have been proposed which
mimic or reproduce the correct relativistic shift in the argument of
the perihelia of the bodies under the gravitational effect of a
central mass \citep{Nob86, satre}. But, as pointed out by
\cite{satre94}, these models  correctly recover the
perihelion motion but fail in reproducing the evolution of the mean anomaly, $M$, point that we will discuss in this paper.

These $r$-dependent secular models have a disadvantage
when used to compute high eccentricity  low perihelion
orbits in constant time-steps algorithms of the type
``advance-impulse-advance", as is the usual case in sympletic
algorithms. In this case, the relativistic perturbation, when
computed as an impulse, generates spurious results near the
perihelion where the relativistic perturbation is stronger and poorly
sampled. This is because the expression for the acceleration has a
term that goes with $1/r^3$ and for low $r$ values this
impulse can be excessively high. For example, in the Solar
System the relativistic perturbation for an asteroid at $r<0.1$ AU
is greater than the gravitational perturbation by Jupiter.
\citet{satre94}, in a sympletic method for planetary integrations, incorporate part of the relativistic correction (\ref{quinn}) in
the ``advance" part of the algorithm but an acceleration proportional
to $1/r^3$ remains as part of the ``impulse".

In this paper we first review the relativistic effects on orbital elements with special attention to
 mean anomaly \citep{rubi,calu,iorio07}. Then, we propose a very simple model that correctly reproduces the evolution of the argument of the perihelion, even for high eccentricity orbits,
 and
which is very useful  for sympletic or, more generally, constant timestep algorithms.

\section{Time Averaged Gauss Planetary Equations}

In order to compute the long term variations in the orbital elements
produced by the perturbations it is a standard procedure to use the
Gauss equations of planetary motion  \citep{Brumbergbook,Murray, Beutler}. As we are interested in the
secular evolution that these corrections generate, we calculate, in
the first place, the Time Averaged Gauss Planetary Equations
(TAGPE). To do so, we have to insert the radial, transverse, and
normal components of the perturbing force ($R$, $T$ and $N$
respectively) into the Gauss equations, evaluate them on an unperturbed
keplerian ellipse, integrate the equations over one orbital
revolution (making use of the change of coordinates
$dt=\frac{r^{2}}{h}df $) and divide them by the orbital period. So
the mean variation rates are the following:

\begin{eqnarray}
\nonumber <\dot{a}> &=& \frac{1}{\sqrt{\mu a}(1-e^{2})\pi} \int_0^{2\pi} \Big[e  R \sin f + T (1 + e \cos f)\Big] r^{2}df  \\
\nonumber <\dot{e}> &=& \frac{1}{2\pi \sqrt{\mu a^{3}}} \int_0^{2\pi} \left[R\sin f +T\left(\cos f + \frac{1}{e} \left(1-\frac{r}{a}\right)\right)\right]r^{2}df \\
\nonumber <\dot{i}> &=& \frac{1}{2\pi (1-e^{2})\sqrt{\mu a^{5}}}\int_0^{2\pi}N \cos (f+\omega) r^{3}df \\
\nonumber <\dot{\Omega}> &=& \frac{1}{2\pi (1-e^{2})\sqrt{\mu a^{5}}\sin i}\int_0^{2\pi} N\sin (f+\omega) r^{3}df \\
\nonumber <\dot{\omega}> &=& \frac{1}{2\pi e\sqrt{\mu a^{3}}}\int_0^{2\pi}\left[-R\cos f + T\sin f \left(\frac{2+e\cos f}{1+e \cos f }\right)\right]r^{2}df - \cos i <\dot{\Omega}>\\
<\dot{M}> &=& \frac{-1}{\pi \sqrt{\mu a^{5}(1-e^{2})}}
\int_0^{2\pi}R r^{3}df - \sqrt{1-e^{2}}(<\dot{\omega}> +
<\dot{\Omega}>\cos i) \label{gaussa}
\end{eqnarray}

where $r= \frac{a(1-e^{2})}{1+ e\cos f}$ and $f$ is the true
anomaly.

\begin{figure}
\includegraphics[width=84mm]{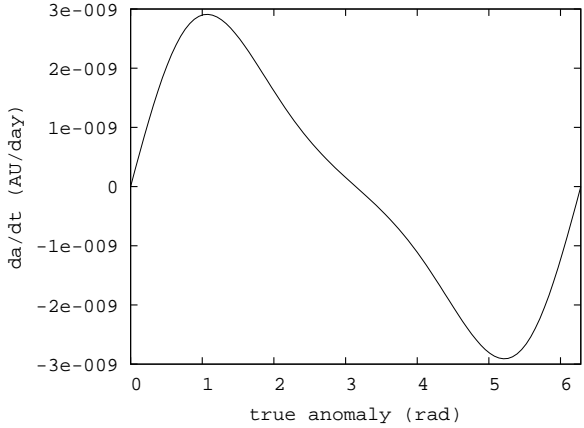}
\caption{Daily variations in semimajor axis generated by Eq. (\ref{quinn}) as deduced from the Gauss equations for planet Mercury. The relativistic
perturbation induces an oscillation in the semimajor axis.}
\label{figure1}
\end{figure}

These expressions are valid assuming the orbital elements remain
constant over an orbital period which is not strictly the case. For
example, there exist very small amplitude periodic variations in the
orbital elements generated by Eq. (\ref{quinn}) \citep{rich} that have negligible
effects on the Gauss equations except for the mean anomaly which is
very sensitive to the variations of $a$. At Fig. \ref{figure1} it is
shown the variation of $da/dt$ in one orbital period for Mercury due
to Eq.(\ref{quinn}). As the semimajor axis is not constant during
one orbital period, the equation for $<\dot{M}>$ should be
considered as the secular relativistic effect on a body with a
semimajor axis which has a mean value given by $a$.

\section{ Relativistic Effects due to the Central Star}

The relativistic acceleration generated by a non rotating central massive body given by Eq. (\ref{quinn})
 can be decomposed in a radial ($R$) and transverse ($T$) components:

\begin{eqnarray}
\nonumber  R &=& \frac{\mu^{2}}{r^{3}c^{2}}\left[2+\frac{r}{a}\left(1+\frac{4e^{2}}{1-e^{2}} \sin^{2}f\right)\right] \\
  T &=& \frac{4\mu^{2}e}{c^{2}}\frac{\sin f}{r^{3}}
\end{eqnarray}
(see also \citealt{dada}).
Substituting in the TAGPE we can compute the secular variations of orbital
elements produced by this acceleration:

\[
<\dot{a}>=<\dot{e}>=<\dot{i}>=<\dot{\Omega}>=0
\]

\begin{equation}
\label{quinnperi}<\dot{\omega}>=\frac{3}{c^{2}(1-e^{2})}\sqrt{\frac{\mu^{3}}{a^{5}}}
\end{equation}

\begin{equation}
\label{quinnanom}<\dot{M}>=\frac{3}{c^{2}}\sqrt{\frac{\mu^{3}}{a^{5}}}\left(2-\frac{5}{\sqrt{1-e^{2}}}\right)
\end{equation}
where units are radians per day.

\begin{figure}
\includegraphics[width=84mm]{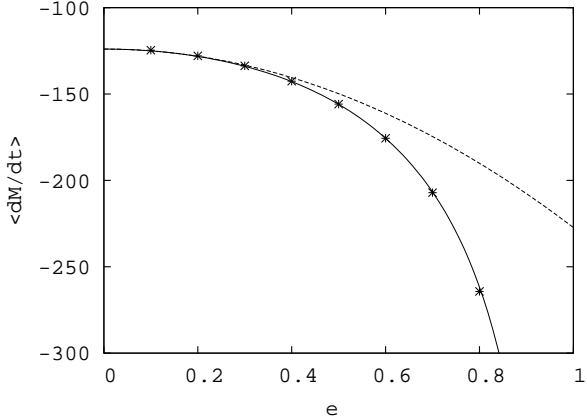}
 \caption{Secular relativistic effect on mean anomaly in arcsec/cy for a particle with Mercury's semimajor axis. Full line:
 formula (\ref{quinnanom}); dashed line: approximation for low eccentricity orbits
 \citep{iorio05b}; dots: numerical results comparing two integrations with and without the
 relativistic perurbation given by Eq. (\ref{quinn}).}
\label{figure2}
\end{figure}

Eq. (\ref{quinnperi}) is the well known relativistic effect on the
argument of the perihelion and, for small eccentricities, Eq.
(\ref{quinnanom}) coincides, for example, with the approximate expressions for
the secular drift in $M$ given by \cite{iorio05b} and with the secular
drift in $M$ generated by the relativistic model for low eccentricities used by \cite{vita}. The exact
formula, valid for all eccentricities, is Eq. (\ref{quinnanom}). We
have checked this performing some numerical integrations
with and without
relativistic effects
using the
packages Mercury \citep{chambers} and Evorb \citep{fer02} of a
system composed by only the Sun and a massless particle with the
semimajor axis of Mercury and varying $e$ from $0$ to $0.9$. (see Fig. \ref{figure2}).
Some caution must be taken in computing $<\dot{M}>$ by means of comparing two numerical integrations.
Because of the oscillations on the semimajor axis
generated by the relativistic correction, a particle integrated with the relativistic model
is a particle that during the integration, on average, had a mean semimajor axis $\bar{a}\neq a_0$
being $a_0$ the initial value. Therefore, to compute the shift on $M$
properly, one must compare the relativistic result with a classical
integration in which the particle has a semimajor axis given by $\bar{a}$, not $a_0$. Otherwise one
would be comparing integrations of particles with different semimajor axes, and
consequently, the obtained $\Delta M = M_{rel} -M_{cla}$  would be meaningless.

\begin{figure}
\includegraphics[width=84mm]{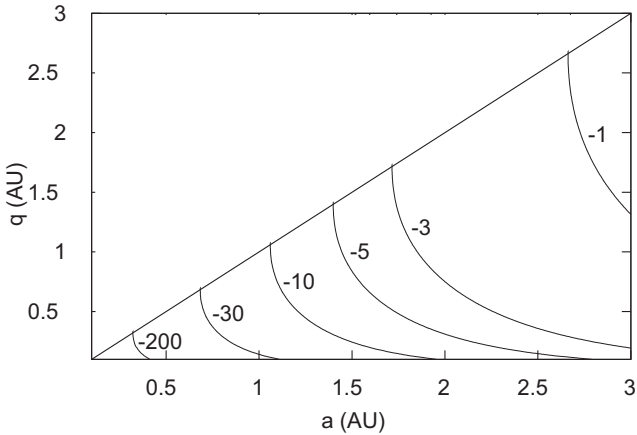}
\caption{Relativistic effect on mean anomaly (arcsec/cy) according to formula (\ref{quinnanom}) for a massless particle in
the gravitational field of the Sun.}
\label{figure3}
\end{figure}

\begin{figure}
\includegraphics[width=84mm]{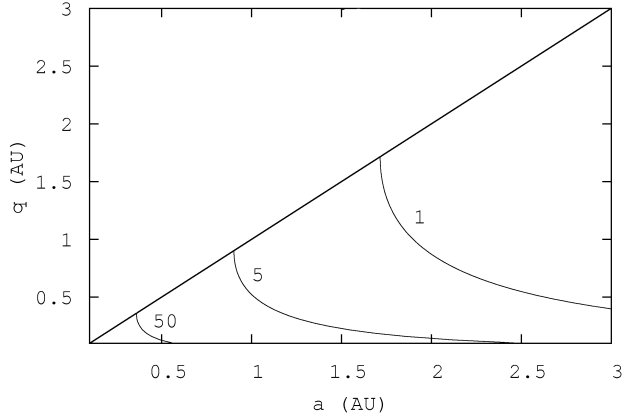}
\caption{Relativistic effect on the argument of the perihelion (arcsec/cy) according to formula (\ref{quinnperi}) for a massless particle in
the gravitational field of the Sun.}
\label{figure4}
\end{figure}

More problems arise when one wants to take into account the drift on mean
anomaly of a specific real body in a numerical integration. The uncertainty $\Delta a$ on the
determination of the semimajor axis of real bodies generates an error in the mean
motion ( $\Delta n \simeq \sqrt{\mu/a^{5}}\Delta a $ ), which produces
an uncertainty of the same order on the drift of mean anomaly.
Setting $\Delta n \approx <\dot{M}>$ and using Eq. (\ref{quinnanom}), one finds that if
\[
\Delta a \gtrsim \frac{3\mu}{c^{2}}(\frac{5}{\sqrt{1-e^{2}}}-2)
\]
then the
uncertainty on the osculating elements generates a secular drift on
$M$ greater than the relativistic effect itself.
 That means that for properly taking into account relativistic effect on $M$
it is necessary to know with high precision, at least of the order
of $10^{-8}$ AU for Solar System's bodies,
the osculating
semimajor axis. That is the case of the terrestrial planets from
Mercury to Mars which have very well defined semimajor axes. On the contrary, the major planets have
uncertainties significantly bigger \citep{pitj}.
The relativistic effect in mean anomaly given by Eq. (\ref{quinnanom}) and checked in our numerical tests  is showed in Fig. \ref{figure3}
as a function of the semimajor axis and perihelion distance. For comparison Fig. \ref{figure4} shows the effect in $\omega$ given by
Eq. (\ref{quinnperi}).

\section{Models that Mimic the Secular Relativistic Effects}

\subsection{Proposed Models}

If we are interested in computing the relativistic effects  due to the Sun
 on a small body, we could just introduce
Eq. (\ref{quinn}) into an integrator. The problem is that this
acceleration depends on both vectors position an velocity of the
particle, so the speed of the integrator may be slowed
down in order to calculate accurately the vectorial products at small
heliocentric distances. Moreover, it is not a simple task to introduce this perturbation
in a sympletic integrator, though it can be done \citep{satre94}.
 In order to overcome this difficulty, some alternative simpler models
have been created in the last two decades.
The relativistic precession of the argument of perihelion is
correctly reproduced defining a radial ($T=0$) acceleration:
\begin{equation}
\label{nob}R=-\frac{6\mu^{2}}{c^{2}r^{3}}
\end{equation}
\citep{Nob86}. Inserting this $R$
in the TAGPE the exact secular drifts generated by Eq. (\ref{quinn}) are
recovered except for mean anomaly.
\cite{satre} added one more term into (\ref{nob}) in order to
account for both $\omega$ and $M$ drifts:
\begin{equation}
\label{saha}R=-\frac{6\mu^{2}}{c^{2}r^{3}} + \frac{3\mu^{2}}{ac^{2}}\left(\frac{4}{\sqrt{1-e^{2}}}-1\right)\frac{1}{r^{2}}
\end{equation}
Now, inserting this perturbation in the TAGPE, Eq. (\ref{quinnperi}) and Eq. (\ref{quinnanom}) are recovered.
This apparent improvement to the original model of \cite{Nob86} is
not so at all
as pointed out by \cite{satre94}.
The point is that given an initial osculating $a_0$ the mean $\bar{a}$ generated by the evolution under Eq. (\ref{quinn})
is different from the mean $\bar{a}_R$ generated by the evolution under Eq. (\ref{saha}). Then, model (\ref{saha}) will not reproduce the secular evolution of an object with the correct $\bar{a}$ but with a different mean semimajor axis given by  $\bar{a}_R$.
In consequence, the model introduces a secular drift in $M$ and no evident progress is done in comparison with Eq. (\ref{nob}).
This problem can be solved taking appropriate initial conditions as we explain in sec. 4.3.

\subsection{Our $r$-independent Models}
The corrections exposed above, though computationally better than Eq. (\ref{quinn})
because the \textbf{v}-dependance is
eliminated; still have the problem that near perihelion the
perturbation can be high enough to introduce numerical errors in
constant time-step integrators.\par
 One way of overcoming this difficulty is to look for a
constant $r$-independent relativistic correction. To accomplish this
task, we went back to Eq. (\ref{quinn}), and assuming that the short
period terms in the orbital elements do not affect the secular
evolution of them, we time averaged the relativistic perturbation produced by a
central body, obtaining:

\[
<\Delta \ddot\textbf{r}>= -\frac{\mu^{2}}{c^{2}a^{3}\sqrt{(1-e^{2})^{3}}}  \mathbf{u_e}
\]

where $\mathbf{u_e}$ is the versor pointing to the pericenter.
The fact that this averaged acceleration is a constant along
$\mathbf{u_e}$, gave us the idea of trying with a constant
perturbation in this direction. In order to obtain the same drift in
$\omega$ as Eq. (\ref{quinnperi}) using the TAGPE it is necessary to introduce a factor 2 and
in terms of $R$ and $T$ can be written as

\begin{eqnarray}
\label{apsides}
 \nonumber
  R &=& -\frac{2\mu^{2}e}{c^{2}a^{3}\sqrt{(1-e^{2})^{3}}}\cos f \\
  T &=& \frac{2\mu^{2}e}{c^{2}a^{3}\sqrt{(1-e^{2})^{3}}}\sin f
\end{eqnarray}

Inserting these expressions into the TAGPE, the elements $a,e,i,\Omega$ have zero variation and the mean variation in the $\omega$ is recovered, but for $M$ we obtain:

\[
<\dot{M}>=-\frac{3(1+e^{2})}{c^{2}}\sqrt{\frac{\mu^{3}}{a^{5}(1-e^{2})^{3}}}
\]
which is different from the real effect given by (\ref{quinnanom}).
Even so, one can also use a simpler constant radial perturbation ($T=0$)  that generates
the expected variation in the argument of perihelion:
\begin{equation}\label{radial} R=\frac{3\mu^{2}}{c^{2}a^{3}\sqrt{(1-e^{2})^{3}}}
\end{equation}
Again, from the TAGPE, the elements $a,e,i,\Omega$ have zero variation and the mean variation in the $\omega$ is recovered but for mean anomaly we obtain a different expression:
\begin{equation}\label{anomrad}<\dot{M}>
=-\frac{9}{c^{2}}\sqrt{\frac{\mu^{3}}{a^{5}(1-e^{2})^{3}}}
\end{equation}
but which is coincident with (\ref{quinnanom}) for $e\longrightarrow 0$.
Therefore, with both constant models the drift in $\omega$ is
recovered but there is a discrepancy for the drift in $M$. This is
irrelevant in numerical simulations because, as we have explained, the predictions for $M$
given by models cannot coincide with the true relativistic
effect if the same set of initial conditions for all models are taken.

The last model that we propose (the constant radial acceleration) is
somehow better than the one along $\mathbf{u_e}$ because is computationally less demanding
and also, in the case
of low eccentricities, both Eq. (\ref{quinnanom}) and Eq. (\ref{anomrad}) give the same result.
Moreover, since model (\ref{radial}) only depends on
$(a,e)$ it is not necessary to evaluate the magnitude of the perturbation in each time step.
The most important point is that
model (\ref{radial}) maintains the precision of the numerical integration even for very small perihelion orbits while
the original Eq. (\ref{quinn}) needs a strong reduction in the step size.

\subsection{On Initial Conditions}

Initial conditions should be used according to the theory they
were determined. For example, the model of \citet{satre} can be used confidently
for generating precise ephemeris only if the initial conditions were obtained
adjusting observations to this model. That is the underlying idea of
the very accurate method for computing ephemeris of low eccentricity orbits proposed by \citet{vita}. But, it is not possible
to follow the exact evolution of the mean anomaly if
there is no consistence between the initial conditions (which they
also should be very precise) and the model used. Then, any of the
models we have analyzed are valid to follow the secular evolution of
an orbit and, in particular, our constant radial perturbation model is
computationally more convenient.

\begin{figure}
\includegraphics[width=84mm]{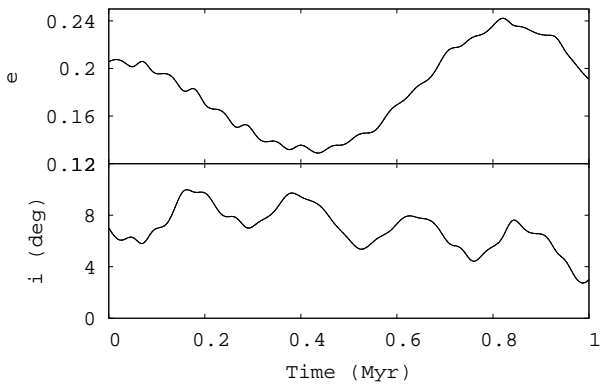}
\caption{Results for planet Mercury from a Solar System's relativistic numerical integration. Evolution of eccentricity and inclination
according to models (\ref{quinn}), (\ref{saha}) and
(\ref{radial}). The numerical results of the all three different models
coincide. A comparison with a non relativistic integration can be found in \citet{Benitez}.}
\label{figure5}
\end{figure}

\begin{figure}
\includegraphics[width=84mm]{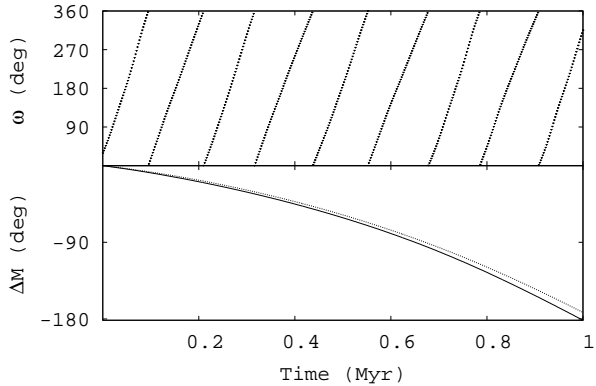}
 \caption{Top:
evolution of Mercury's argument of perihelion according to models
(\ref{quinn}), (\ref{saha}) and (\ref{radial}). Al three models are coincident. Bottom: differences
on Mercury's mean anomaly between model (\ref{radial}) and
(\ref{quinn}) (full line) and between model (\ref{saha}) and
(\ref{quinn}) (dotted line). In spite of these differences the secular evolution is the same.}
\label{figure6}
\end{figure}

For illustrative purposes, in order to compare the variations of the
orbital elements produced by the different models given by  Eq.
(\ref{quinn}), Eq. (\ref{saha}) and Eq. (\ref{radial}), we integrated
numerically our planetary system using Evorb  for one million years.
Results for planet Mercury are shown in Fig. \ref{figure5} and Fig.
\ref{figure6}. These figures corroborate what we have already stated
by means of the Gauss equations: the secular relativistic evolution
of all orbital elements is very well reproduced except for mean
anomaly (a comparison with the classic evolution can be found in
\citealt{Benitez}). And even with strong differences in $M$ (Fig.
\ref{figure6} bottom), the orbital evolution of the different models
is almost undistinguishable. For example, at the end of the
integration, differences of the order of $10^{-6}$ AU in $a$,
$10^{-5}$ in $e$, $10^{-3}$ degrees on $\omega$, $10^{-4}$ degrees
on $\Omega$, and even lower differences for $i$ were obtained.
Note also that the numerical deviations on $M$ with respect to model
(\ref{quinn}) in Fig. \ref{figure6} are mostly a consequence of
using the same initial conditions for integrations that deal with
different models. In the case of model (\ref{saha}), the drift on $M$
will disappear if initial conditions consistent with this model
are taken. In the case of model (\ref{radial}) some drift will persist.

\section{Conclusions}

By means of numerical integrations we confirmed the predicted secular drift in mean anomaly given by formula (\ref{quinnanom}), which is valid
for all eccentricities and was obtained introducing the relativistic effect due to the Sun (Eq. \ref{quinn}) into
the time averaged Gauss' planetary equations.
Several models are able to reproduce the drift in $\omega$ but for high eccentricity orbits only the model by \citet{satre} can follow the exact
secular drift in $M$ and only if appropriate initial conditions are taken.
For secular evolution studies of numerous populations by means of numerical integrations, if no precise ephemeris are required, our constant radial independent model is very convenient, specially if low perihelion orbits are involved.
For accurate
ephemeris computations of real bodies, the original formula (\ref{quinn}) or even the full relativistic N-body one should be used with a control of the integrator's precision
in the case of low perihelion orbits.

\section*{Acknowledgments}
J. Venturini acknowledges a scholarship from PEDECIBA. This work was done as part of
the Project ``Caracterizaci\'{o}n de las poblaciones de cuerpos menores del sistema solar" (ANII). We acknowledge
comments by anonymous referees.

\label{lastpage}

\end{document}